\documentclass{ieeeaccess}
\usepackage{cite}
\usepackage{amsmath,amssymb,amsfonts}
\usepackage{algorithmic}
\usepackage{tabularx}
\usepackage{setspace}
\usepackage{tablefootnote}
\usepackage{makecell}
\usepackage{amssymb}
\usepackage{amsmath}
\usepackage{hyperref}
\usepackage{array}
\usepackage{geometry}
\usepackage{booktabs} % For better table rules
\usepackage{multirow}% For multirow cells
\usepackage{geometry}
\usepackage{array}
\usepackage{subfigure}
\usepackage{hyperref}
\usepackage{setspace}
\hypersetup{
    colorlinks=false,
    linkcolor=false,
    filecolor=false,      
    urlcolor=false,
    }
\usepackage{textcomp}
\def\BibTeX{{\rm B\kern-.05em{\sc i\kern-.025em b}\kern-.08em
    T\kern-.1667em\lower.7ex\hbox{E}\kern-.125emX}}

\usepackage{colortbl}
\usepackage[table,xcdraw]{xcolor}
\usepackage{setspace}
\usepackage{array}
\usepackage{enumitem}
\hypersetup{
    colorlinks=false,
    linkcolor=false,
    filecolor=false,      
    urlcolor=false,
    }
\usepackage{textcomp}
\def\BibTeX{{\rm B\kern-.05em{\sc i\kern-.025em b}\kern-.08em
    T\kern-.1667em\lower.7ex\hbox{E}\kern-.125emX}}

    \usepackage{geometry}
\usepackage{booktabs} % For better table rules
\usepackage{multirow}% For multirow cells
\usepackage{geometry}
\usepackage{array}

\usepackage{setspace}
\hypersetup{
    colorlinks=false,
    linkcolor=false,
    filecolor=false,      
    urlcolor=false,
    }
\usepackage{textcomp}
\def\BibTeX{{\rm B\kern-.05em{\sc i\kern-.025em b}\kern-.08em
    T\kern-.1667em\lower.7ex\hbox{E}\kern-.125emX}}

\usepackage{colortbl}
\usepackage[table,xcdraw]{xcolor}

\usepackage[demo]{graphicx}

\begin{document}
\history{Date of publication xxxx 00, 0000, date of current version xxxx 00, 0000.}
\doi{10.1109/ACCESS.2023.0322000}

\title{User Identity Linkage on Social Networks: A Review of Modern Techniques and Applications}
\author{\uppercase{Caterina Senette}\authorrefmark{1},
\uppercase{Marco Siino}\authorrefmark{2}, and \uppercase {Maurizio Tesconi}\authorrefmark{1}
}

\address[1]{Institute of Informatics and Telematics, National Research Council of Italy (IIT-CNR) (e-mail: (caterina.senette, Maurizio.tesconi)@iit.cnr.it)}
\address[2]{Dipartimento di Ingegneria Elettrica
Elettronica e Informatica
University of Catania
Catania, Italy (e-mail: marco.siino@unipa.it)}
%\address[3]{Electrical Engineering Department, University of Colorado, Boulder, CO 80309 USA}

%\tfootnote{This work was supported in part by Project SERICS (PE00000014) under the NRRP MUR program funded by the EU – NGEU}
\tfootnote{\textcolor{red}{This work has been submitted to the IEEE for possible publication. Copyright may be transferred without notice, after which this version may no longer be accessible.}}

\markboth
{Senette \headeretal: User Identity Linkage on Social Networks}
{Senette \headeretal: User Identity Linkage on Social Networks}

\corresp{Corresponding author: Caterina Senette (e-mail: caterina.senette@iit.cnr.it).}

\begin{abstract}
In an Online Social Network (OSN), users can create a unique public persona by crafting a user identity that may encompass profile details, content, and network-related information. As a result, a relevant task of interest is related to the ability to link identities across different OSNs.
Linking users across social networks can have multiple implications in
several contexts both at the individual level and at the group level. At the
individual level, the main interest in linking the same identity across social networks is to enable a better knowledge of each user. At the group level, linking user identities through different OSNs helps in predicting user behaviors, network dynamics, information diffusion, and migration phenomena across social media.
The process of tying together user accounts on different OSNs is challenging and has attracted more and more research attention in the last fifteen years. The purpose of this work is to provide a comprehensive review of recent studies (from 2016 to the present) on User Identity Linkage (UIL) methods across online social networks. This review aims to offer guidance for other researchers in the field by outlining the main problem formulations, the different feature extraction strategies, algorithms, machine learning models, datasets, and evaluation metrics proposed by researchers working in this area. The proposed overview takes a pragmatic perspective to highlight the concrete possibilities for accomplishing this task depending on the type of available data.
\end{abstract}


\titlepgskip=-21pt

\maketitle

\section{Introduction}
\label{sec:introduction}
\PARstart{E}{VERYONE's} social life has been changed by the recent growth of social network services of all kinds, which make it easier and more enjoyable than ever to share a variety of information (e.g., microblogs, images, videos, reviews, location check-ins). How to use this large amount of social data for improved business intelligence is undoubtedly the biggest and most fascinating topic facing all firms. People are particularly concerned with understanding each individual user more effectively, given the vast amount of social data now available. Unfortunately, a user’s social scene information is fragmented, unreliable, and disruptive. Due to the wide variety of services offered by online social networks (OSNs), it seems natural for users to register for accounts (also known as user identities) on many OSNs. Having accounts (also known as user IDs) on several OSNs has grown in popularity. 
According to a 2023 statistic\footnote{\url{https://wearesocial.com/it/blog/2023/01/digital-2023-i-dati-globali/ }(2024-09-10)}, among 50 nations with internet users aged between 16 to 64, Japan had the lowest overall number of social media accounts at 3.5 per user, while India had the highest at 9, the average number around the world is 7.2 accounts per user.

User Identity Linkage (UIL) refers to the process of linking or matching user identities across different online platforms or social networks by analyzing the similarities in their profiles, behaviors, or activities. The problem is also known as  Social Identity Linkage \cite{liu2014hydra}, User Identity Resolution  \cite{bartunov2012joint}, Social Network Reconciliation  \cite{korula2013efficient},  Profile Linkage  \cite{zhang2014online}, Anchor Link Prediction  \cite{kong2013inferring}. It is used to consolidate user information from multiple sources, providing a comprehensive view of an individual across platforms. It is commonly employed in domains like personalized recommendations, cross-platform marketing, and more recently, in cyber intelligence to detect malicious actors.

Linking users across social networks can have multiple implications in several contexts, both at the individual level and at the group level. At the individual level, the main interest in linking the same identity across social networks is to enable a better knowledge of each user by aggregating all the information collected from each social network platform. A more comprehensive frame of a single user simplifies strategies for cross-system personalization that in turn could be used to build user-adaptive systems able to trigger external recommendation systems and to provide personalized services in e-commerce, tourism, travel planning, and so on \cite{carmagnola2009user, deng2013personalized}. 
By reasoning at the group level, linking user identities through different OSNs helps in predicting user behaviors, network dynamics, and information diffusion other than understanding migration phenomena across social media that in turn could be beneficial for social media site platforms to generate revenue from suggested advertising and to grow their base with the ultimate goal to improve marketing outcomes \cite{kumar2011understanding, zafarani2009connecting}.
The use of UIL techniques is also crucial in the field of cyber intelligence, as it enables the identification and tracking of suspicious activities across different social platforms, facilitating the detection of malicious actors and the prevention of disinformation campaigns or coordinated attacks \cite{cinelli2022}.

For all these reasons, User Identity Linkage has become a trending topic and has attracted more and more research attention. It is not easy to take advantage of the numerous chances that people with profiles on several social media platforms present. Linking users’ accounts across various online social networks provides all the aforementioned opportunities; however, the process of tying together user accounts on different OSNs is challenging. 
The main reasons are:
(i) For the same person in the real world, user identity information on different online social networking sites can vary greatly; (ii) Online social networking data is vast, noisy, imperfect, and largely unstructured. 
Any single social network service can only provide a limited view of a user from a certain perspective due to limitations imposed by the features and design of each service. An otherwise disjointed user profile would be enriched by cross-platform user linking, allowing for a comprehensive grasp of a user’s interests and behavioural patterns. The information posted by users on social media platforms may be incorrect, contradictory, incomplete, and deceptive for a variety of reasons. The consistency of user information can be improved by cross-referencing across several platforms. 

In addition to the aforementioned challenges inherent to the task, the more critical issue lies in the scarcity of publicly available datasets, which limits the ability to train supervised systems and verify experimental results. This lack of comprehensive datasets hinders progress in effectively linking user accounts across platforms.

Although social networks rise and fall, real-world consumers continue to use them and simply switch to newer ones. Linking user identities enables the integration of important user data from platforms that have over time declined in popularity or even been abandoned.

Along with its benefits, the use of UIL also introduces considerable risks, especially regarding privacy. It enables widespread user tracking across platforms, potentially leading to unauthorized profiling, data exploitation, and breaches of user consent.

A portion of the literature on the topic has been summarized by a survey dated 2016 \cite{shu2017user} which presents a unified framework for UIL task consisting of two phases, feature extraction and model construction. Moreover, the authors summarize different aspects of feature extraction and model construction techniques and discuss different datasets and evaluation metrics proposed by existing approaches. 

Comparing state of the art prior to 2016 \cite{shu2017user} with more recent works (up to 2024) permits us to provide an overall look at the UIL in terms of the following aspects: (i) The rise of new problem formulations of the User Identity Linkage problem; (ii) New methods used to extract and represent features from social networks; (iii) Up-to-date AI models built to address the UIL task; (iv) Novel algorithms introduced; (v) Deep focus on data collection and concrete availability of datasets.
\subsection{Research Questions and Contribution}
It is worth noting that, in the last decade, there have been many changes in this research field which can be summarized as follows: (i) The faster growth of social networks and the higher diversity among them; (ii) The increased attention to privacy, which effectively limits the real availability of individual data thus slanting strategies towards creating detailed profiles based on user online activity rather than relying on disclosed personal data; (iii) Computational methods increasingly oriented towards deep learning, nowadays considered as a core technology of today's Fourth Industrial Revolution \cite {sarker2021deep}.

We believe that all these elements would require an update of the state of the art, so, the purpose of the current work is to provide a comprehensive review of very recent studies (from 2016 to date) of User Identity Linkage methods across online social networks with the intention to give guidance for other researchers working in the field in light with the current possibilities to accomplish the UIL task.  

To this end, we will offer an alternative perspective that approaches the problem from a pragmatic point of view expressed by the following research questions (RQs): 
\begin{itemize}
\item{(\textbf{RQ1})\textit {What are the current prevailing problem formulations, methodologies, and techniques used in User Identity Linkage across social networks?}}
\item{(\textbf{RQ2})\textit {What performances do they currently guarantee?} }
\item{(\textbf{RQ3})\textit {What issues are still open in this field?}}
\end{itemize}

By answering these research questions we provide the following contributions:
\begin{itemize}
\item{An updated overview of the body of knowledge on UIL in order to fill the chronological gap with respect to previous literature reviews and to identify the aspects in which the major innovations have been introduced.}
\item{A functional approach to the task with the intention to explore and give some guidance for more practical problem settings in User Identity Linkage across social networks.}
\item{A useful collection of the datasets used for UIL task, not built ad hoc for a single experiment and shared in several research works. 
Each is provided with a reference.}
\end{itemize}
\subsection{Roadmap}

The paper has the following structure. 
After the Introduction, in Section \ref{sec:method} we describe explicit and rigorous criteria used to identify, critically evaluate, and synthesize all the available works on recent literature. In Section \ref{sec:problem} the problem of User Identity Linkage is narrowed down to two possible formulations (the most common ones). Here we introduce our conceptual framework. In Section \ref{Section:Net-Al} and \ref{Section:Class}, we present all the state-of-the-art solutions to perform UIL summarizing them guided by our framework. When appropriate, collected papers are further cataloged based on the category of data used to accomplish the UIL task. Specifically, for each single data type or group of data, we explore all the research works illustrating different feature extraction strategies, different algorithms, and different machine learning models proposed. In Section \ref{sec:evalu} we explore evaluation metrics used in each of the two UIL formulations. In Section \ref{sec:dataset} we present a detailed catalog of all the datasets proposed in the recent literature. Finally, in Section \ref{sec:open} we examine the main challenges still open, and, in Section \ref{sec:conclusion} we draw a conclusion for the paper.

\section{Method}
\label{sec:method}
Candidate papers for inclusion in this review were gathered through four steps.  

As a first step, we conducted a search in Google Scholar and Microsoft Academic which are the most used academic search engines to collect the knowledge base about this topic. We carried out the search in December 2023 and again in June 2024 looking for any potential new entry and using search terms covering variations on "User Identity Linkage". Specifically, we used six different strings as key search: (a) \textit{User Identity Linkage across social networks}; (b) \textit{user accounts linkage across social networks}; (c) \textit{social network alignment}; (d) \textit{network alignment}; (e) \textit{user profile matching across social networks}; (f) \textit{reconciliation across social networks}. 
These searches yielded a total of 102 results which were downloaded.  Among these, we identified only two documents representing scientific reviews that synthesize and integrate knowledge about the topic \cite{shu2017user, kaushal2020systematic}. The work of Shu et al. \cite{shu2017user}, dated 2016, motivates the need for a new update on the topic covering the last eight years, the second work \cite{kaushal2020systematic} is a doctoral dissertation dated 2020 which focused on the study of conventional machine learning based approaches and more recent graph representation based approaches. Therefore the latter contribution lacks a comprehensive guide on the topic. 

As a second step, a careful check of titles and abstracts reduced this to a list of 95 publications selected as relevant for closer attention and consideration for inclusion in the review. 

As a third step, key papers from the list above were identified, and, for those dated after 2016, hand-searches were conducted in their reference sections, looking for related papers respecting inclusion criteria. 
Key paper citations were collected through ad-hoc software (Mendeley). 

As a fourth step, for those papers that released an associated dataset, we identified and, when needed, contacted corresponding authors asking them to share their data repository. This was extremely helpful especially to verify whether the data model underlying theoretical approaches to User Identity Linkage described in their work corresponds exactly to the data actually available. 
As shown in Table \ref{table:dataset} we identified 16 datasets used for the UIL task in more than one research work.

A final list of 85 contributions reporting novelties on the User Identity Linkage task found via these methods is included in the current work and associated contents are detailed in the next sections of this survey.
  \begin{figure*}[t!]
  \centering
  \includegraphics[width=2.0\columnwidth]{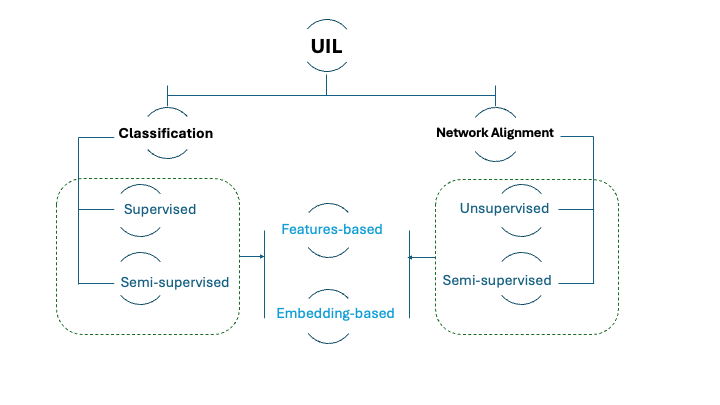}            
  \label{fig:framework}
   \caption{Conceptual Framework}
  \end{figure*}

\section{Problem formulation and general Approach}
\label{sec:problem}
The number of possible problem formulations for user-identity linkage across social networks can vary based on several factors. The final goal might include identifying duplicate accounts, merging user profiles, or enhancing recommendation systems. Formulations can also vary depending on the scope, determining whether the analysis pertains to a single social network or spans multiple ones. Furthermore, the context of the linkage, the desired outcomes, and the methods employed are crucial elements that shape the problem formulation.

Focusing on the objective, the User Identity Linkage (UIL) task can be primarily formulated as a \textit{classification} task or as a \textit{network alignment} task. 

When the objective of UIL is to determine whether two profiles from different networks belong to the same individual, the problem is framed as a classification task. This formulation is primarily addressed using supervised (and semi-supervised) method. These methods leverage labeled data, such as Pre-aligned user Pairs or supervisory anchor pairs (SAPs), to train predictive models. The goal is to learn discriminative features that enable the prediction of whether two profiles represent the same user. This approach is more straightforward and tractable compared to network alignment. However, the success of supervised methods heavily depends on the availability and quality of SAPs. Despite their scarcity in real-world scenarios, these pairs are crucial for training accurate and reliable models. Results achieved by some authors \cite{guo2020user,qiao2020dual} indicate that even with a minimal amount of carefully selected SAPs, the overall performance of the models is significantly boosted.

Alternatively, when the objective of UIL is to align nodes from different social networks based on their structural attributes without relying on labeled data, the problem is conceptualized as a network alignment challenge. Unsupervised or semi-supervised approaches are employed to achieve this goal. Network alignment techniques aim to map the entire structure of one network onto another, aligning nodes based on structural similarities. This thorough alignment often results in problems that are highly complex and difficult to solve optimally. The challenge becomes particularly pronounced in networks that are either very dense or exceptionally large \cite{zhang2017link,liu2017deep}.

The narrative perspective used herein in this survey is guided by this conceptual framework as shown in Figure \ref{fig:framework}

\subsection{Features-based VS embedding-based approaches}

As highlighted in Figure \ref{fig:framework} both problem formulations share two primary sub-approaches, feature-based and embedding-based strategies used to identify and match user accounts across social networks.
Features are measurable attributes or properties derived from the raw data and processed/transformed to be used in machine learning models or other analytical processes. We provide a categorization of them in Table \ref{tab:profile_matching_features}.

\begin{figure}[ht]
  %\centering
  \includegraphics[width=1.0\columnwidth]{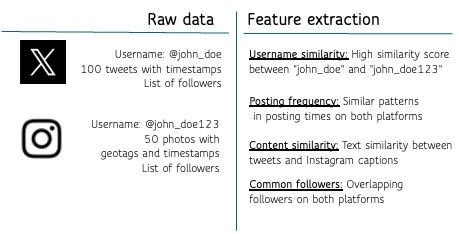}            
  \label{fig:feat_ext}
  \caption{From Raw-Data to Features: an example of UIL across X (Twitter) and Instagram} 
\end{figure}

Additionally, Figure \ref{fig:feat_ext} shows an example of feature extraction from row data in the scenario of UIL across X (Twitter) and Instagram.

Features-based methods traditionally involve manually designing features to capture similarities and differences between user profiles across networks. These features might include profile attributes, interaction patterns, and network connections. This approach explicitly defines and extracts the features from raw data, making them interpretable but potentially limited by the quality and relevance of the chosen features.

On the other hand, embedding-based methods (particularly recent unsupervised and semi-supervised techniques) \cite{mikolov2013distributed, mikolov2013efficient}, utilize graph embedding techniques to automatically learn low-dimensional latent vectors (embeddings) that capture the structural properties of network nodes. This latent space representation allows nodes to be represented in a continuous space where similar nodes are closer together, enhancing the methods' flexibility and ability to capture complex data patterns. Embedding-based methods generally outperform traditional feature-based approaches \cite{shu2017user, liu2020user}, leveraging deep learning techniques to handle large-scale data and capture intricate structural relationships more effectively.

\begin{figure*}%
 
 \label{Feat_VS_Embed}
    \centering
\subfigure{{\includegraphics[width=7cm]{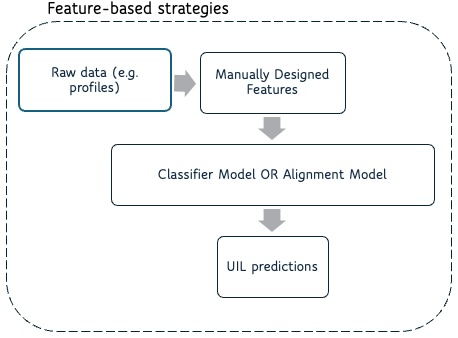} }}%
    \qquad
\subfigure{{\includegraphics[width=7cm]{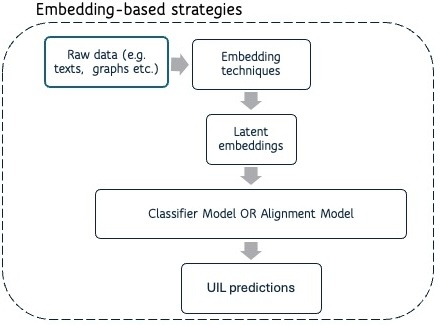} }}%
   
    \label{fig:example}%
\caption{Feature-based VS Embedding-based approach}%
\end{figure*}

Figure \ref{Feat_VS_Embed} provides one flow diagram for each strategy. Additionally, Table \ref{tab:comparison} summarizes the aspects to consider when comparing feature-based methods with embedding-based methods in the two main User Identity Linkage problem formulations identified in this survey.

It is worth noting that while feature-based and embedding-based strategies have their own strengths and weaknesses, they are not mutually exclusive and can be integrated into more robust UIL systems. For example, feature-based approaches can provide interpretable insights that help guide the design of embedding models. Conversely, embeddings can be used to enhance feature-based models by providing additional learned features \cite{xu2020deep, zhang2018combining, zhang2020cross}.

%%PROVA FIGURE
%%FINE PROVA

 %% NUOVA TABELLA
\begin{table*}[t!]
\centering
\caption{Features for Social Media Profile Matching.}
\small % Adjust font size to small
\setstretch{1.1} % Adjust line spacing

\begin{tabular}{@{} p{0.25\linewidth} p{0.6\linewidth} @{}}
\toprule
\textbf{\footnotesize Feature Category} & \textbf{\footnotesize Details} \\
\hline
\midrule
\textbf{\footnotesize Direct Matching Features} & 
Identical Identifiers: Username, Email address, Phone number, Social Security number \\
& Account IDs or handles \\
\midrule
\textbf{\footnotesize Behavioral Features} & 
Posting Frequency: Number of posts, Time between posts \\
& Interaction Patterns: Number of likes, Number of comments, Number of shares, Types of interactions (e.g., reactions, emojis used) \\
& Activity Timestamps: Time of day for activity, Days of the week for activity \\
& Engagement Levels: Total engagement (likes, comments, shares), Engagement per post \\
\midrule
\textbf{\footnotesize Attribute Similarity Features} & 
Name Similarity: Levenshtein distance between names, Jaccard similarity of names \\
& Age or Birthdate Comparison: Age difference, Birth year comparison \\
& Education or Workplace: Comparison of educational institutions, Comparison of workplace names \\
& Gender or Pronouns: Matching gender information, Pronoun usage in profile descriptions \\
\midrule
\textbf{\footnotesize Graph-Based Features} & 
Shared Connections: Mutual friends or followers, Shared groups or communities \\
& Social Network Graph Metrics: Degree centrality, Betweenness centrality, Clustering coefficients \\
\midrule
\textbf{\footnotesize Textual Features} & 
Profile Description Similarity: TF-IDF similarity of profile descriptions, Word embeddings (e.g., Word2Vec) similarity \\
& Post Content Similarity: TF-IDF similarity of post content, Cosine similarity of post embeddings \\
\midrule
\textbf{\footnotesize Metadata Features} & 
Interests or Hobbies: Similarity of interests, Comparison of liked pages or groups \\
& Profile Completeness: Completeness of profile information, Presence of specific attributes (e.g., "interests" section) \\
\midrule
\textbf{\footnotesize Image-Based Features} & 
Profile Picture Similarity: Image hashing similarity, Deep learning feature similarity (e.g., CNN features) \\
\midrule
\textbf{\footnotesize Spatio-Temporal Features} & 
Account Creation Date Difference, Last Active Timestamp Comparison, Time Since Last Activity \\
& Location Proximity: Distance between reported locations, Common locations or regions \\
\bottomrule
\end{tabular}
\label{tab:profile_matching_features}
\end{table*}

In the next sections \ref{Section:Net-Al} and \ref{Section:Class}, we will describe the relevant literature categorizing all the approaches proposed by scholars based on the type of UIL problem formulation as introduced in our conceptual framework. 
This overview will answer the first research question (\textbf{RQ1}). 

A synthesis of algorithms cited in this section is shown in Table \ref{tab:algo}.

%questa sarebbe la tabella che fa
%un sunto del confront fra I 2 approcci - da mettere dopo le figurine relative agli approcci singoli
%PROVA TABELLA NEW (deve stare nella sezione III)
\begin{table*}[h!]
\setlength{\tabcolsep}{4pt}
\setlength{\arrayrulewidth}{0.5mm}
\renewcommand{\arraystretch}{1.5}
\centering
\caption{Comparison of Features-Based vs. Embedding-Based Methods in User Identity Linkage Problem}
\begin{tabular}{>{\raggedright\arraybackslash}p{2.5cm}|>{\raggedright\arraybackslash}p{2.5cm}>{\raggedright\arraybackslash}p{2.5cm}|>{\columncolor{blue!20}}>{\raggedright\arraybackslash}p{3.cm}>{\columncolor{blue!20}}>{\raggedright\arraybackslash}p{3.cm}}
\hline
\textbf{\#\#\#} & \textbf{Features-Based (Sup)} & \textbf{Features-Based (UNSup/Semi-Sup)} & \textbf{Embedding-Based (Sup)} & \textbf{Embedding-Based (UNSup/Semi-Sup)} \\ \hline \hline
%\hline
\textbf{Problem Type} & Classification & Network Alignment & Classification & Network Alignment \\ \hline
\textbf{Feature Extraction} & Attribute and structural features & Attribute and structural features & Low-dimensional embeddings & Low-dimensional embeddings \\ \hline
\textbf{Training Data} & Pre-aligned user pairs (SAPs) & None & Pre-aligned user pairs (SAPs) & None or limited SAPs \\ \hline
\textbf{Model} & Binary classifiers (e.g., SVM, logistic regression) & Graph matching, clustering & Neural networks, advanced classifiers (BERT, MLP, GPT, etc.) & Graph neural networks, embedding matching \\ \hline
\textbf{Evaluation Metrics} & Accuracy, precision, recall, F1-score & Clustering quality, structural preservation & Accuracy, precision, recall, F1-score & Alignment accuracy, MAP, etc.\\ \hline
\textbf{Scalability} & Limited by feature complexity and classifier & Limited by network size and density & Can handle large datasets with appropriate embedding & Effective for large, complex networks \\ \hline
\end{tabular}
\label{tab:comparison}
\end{table*}

\section{UIL as a Network Alignment Problem}
\label{Section:Net-Al}
As already said, when addressing the User Identity Linkage (UIL) problem as a network alignment task the process focuses on structural properties and attributes similarities between nodes without using labeled data. 

\subsection{Feature-based strategies}
Table \ref{tab:Net-Al-Features} provides a general scheme detailing how this type of task works given a set of input data.
It begins with feature extraction, then, algorithms like graph matching or clustering are employed to align user accounts across different networks, aiming to identify the optimal alignment based on these features. Evaluation of these methods typically involves assessing how well the network structure is preserved and using indirect metrics, such as clustering quality, to gauge the effectiveness of the alignment.

An unsupervised approach called Friend Relationship-based User Identification algorithm without Prior (FRUI-P) knowledge is proposed in \cite{zhou2017structure}, after observing that friend relationships are trustworthy and consistent across distinct SNs. The FRUI-P evaluates the similarities of all the possible identical users between two OSNs after extracting the friend feature of each user in an SN into a friend feature vector. The users are then identified using a one-to-one map approach that takes into account their commonalities.

In \cite{zhong2018colink} the authors proposed the CoLink framework (semi-supervised) that utilized a co-training algorithm applying two distinct models: an attribute-based model, and a relationship-based model. Both models are designed to perform binary classification, determining whether a given pair of users is positive (linked) or negative (unlinked). The co-training algorithm iteratively enhances the performance of these two models. In each iteration, both models are retrained using the set of linked pairs, defined as \textit{S}. The co-training algorithm must start with a small seed set of linked user pairs. Then the seed set is generated using specially created rules, or "seed rules". The attribute-based model employs sequence-to-sequence learning to handle attribute alignment, while the relationship-based model uses social connections. Despite employing sophisticated techniques, this approach fundamentally relies on manually designed features (attributes and relationships).

%%TAB NETWORK_ALIGNMENT-FEATURE-BASED
\begin{table*}[h!]
\caption{UIL as a Network Alignment problem Features-based}
\centering
\begin{tabular}{|p{13cm}|}
\hline
\textbf{UIL as a Network Alignment problem Features-based} \\ \hline
\textbf{GIVEN:} \\
\begin{itemize}
  \item Two or more social network platforms denoted as $P_1, P_2, \ldots, P_n$.
  \item Each platform $P_i$ contains a set of user accounts denoted as $U_i = \{u_{i1}, u_{i2}, \ldots, u_{im_i}\}$.
  \item Each user $u_{ij}$ has associated attributes or features, such as:
  \begin{itemize}
    \item Username, email address, or other identifiers
    \item Profile information (e.g., name, age, location)
    \item Behavioral patterns (e.g., posting frequency, types of interactions)
    \item Metadata (e.g., interests, groups, friends)
  \end{itemize}
\end{itemize}

\textbf{TASK:} \\
\textbf{Problem Definition:} \\
\begin{itemize}
  \item Define the task as aligning user accounts from different platforms based on their structural and attribute similarities.
\end{itemize}

\textbf{Formalization:} \\
\begin{itemize}
  \item Let $U_i$ and $U_j$ be the sets of user accounts on platforms $P_i$ and $P_j$ respectively.
  \item For a given pair of user accounts $u_{ik} \in U_i$ and $u_{jl} \in U_j$, define the alignment task as finding the best matching based on feature similarities.
  \item The task can be denoted as finding a mapping function $f : U_i \times U_j \to [0,1]$, where a higher value indicates a higher likelihood of the accounts being the same user.
\end{itemize}

\textbf{Feature Representation:} \\
\begin{itemize}
  \item Define a set of features $X_{ik,jl}$ representing the similarity or dissimilarity between the attributes of user accounts $u_{ik}$ and $u_{jl}$.
  \item Features can include:
  \begin{itemize}
    \item Direct matching features (e.g., identical identifiers)
    \item Behavioral features (e.g., posting frequency, interactions)
    \item Attribute similarity features (e.g., name similarity, location proximity)
    \item Graph-based features (e.g., shared friends, groups)
    \item Textual features (e.g., similarity of profile descriptions, posts)
  \end{itemize}
\end{itemize}

\textbf{Model Selection and Evaluation:} \\
\begin{itemize}
  \item Use algorithms such as graph matching or clustering to find the optimal alignment based on the features.
  \item Evaluate the performance of the alignment using metrics such as structural preservation and clustering quality.
  \item Implement the chosen alignment algorithm and test it on a dataset of user pairs $(u_{ik}, u_{jl})$ with corresponding similarity scores.
  \item Use the alignment algorithm to predict the likelihood of a pair of user accounts representing the same individual for new, unseen data.
\end{itemize} \\ \hline
\end{tabular}
\label{tab:Net-Al-Features}
\end{table*}

\subsection{embedding-based strategies}
Table \ref{tab:Net-Al-Embeddings} depicts an outline of how this type of task is performed using a given set of input data.
The process begins with embedding the network data, transforming nodes and their connections into dense vector representations that capture both local and global structural information. Advanced algorithms, such as graph neural networks or matrix factorization techniques, are then employed to align these embeddings across different networks. The goal is to identify the optimal alignment by comparing these learned embeddings. Evaluation of these methods typically involves assessing how well the network's structural properties are preserved in the embeddings and using metrics like alignment accuracy and embedding quality to gauge the effectiveness of the alignment.

\subsubsection{Unsupervised methods}

\begin{table*}[h!]
\caption{UIL as a Network Alignment problem Embedding-based}
\centering
\begin{tabular}{|p{13cm}|}
\hline
\textbf{UIL as a Network Alignment problem Embedding-based} \\ \hline
\textbf{GIVEN:} \\
\begin{itemize}
  \item Two or more social network platforms denoted as $P_1, P_2, \ldots, P_n$.
  \item Each platform $P_i$ contains a set of user accounts denoted as $U_i = \{u_{i1}, u_{i2}, \ldots, u_{im_i}\}$.
  \item Each user $u_{ij}$ has associated attributes or features, such as:
  \begin{itemize}
    \item Username, email address, or other identifiers
    \item Profile information (e.g., name, age, location)
    \item Behavioral patterns (e.g., posting frequency, types of interactions)
    \item Metadata (e.g., interests, groups, friends)
  \end{itemize}
\end{itemize}

\textbf{TASK:} \\
\textbf{Problem Definition:} \\
\begin{itemize}
  \item Define the task as embedding user accounts from different platforms into a common latent space and aligning them based on their embeddings.
\end{itemize}

\textbf{Formalization:} \\
\begin{itemize}
  \item Let $U_i$ and $U_j$ be the sets of user accounts on platforms $P_i$ and $P_j$ respectively.
  \item For a given pair of user accounts $u_{ik} \in U_i$ and $u_{jl} \in U_j$, define the alignment task as finding the best matching based on their embeddings.
  \item The task can be denoted as finding a mapping function $f : (u_{ik}, u_{jl}) \to [0,1]$ in the latent space, where a higher value indicates a higher likelihood of the accounts being the same user.
\end{itemize}

\textbf{Feature Representation:} \\
\begin{itemize}
  \item Define a set of embeddings $E_{ik}$ and $E_{jl}$ for user accounts $u_{ik}$ and $u_{jl}$ respectively, representing their positions in the latent space.
  \item Embeddings can be derived from:
  \begin{itemize}
    \item Structural features (e.g., network topology)
    \item Attribute features (e.g., profile information)
    \item Behavioral features (e.g., interaction patterns)
    \item Combined features (e.g., joint representation of structural and attribute features)
  \end{itemize}
\end{itemize}

\textbf{Model Selection and Evaluation:} \\
\begin{itemize}
  \item Use embedding algorithms such as graph embeddings, node2vec, or GCNs to learn the embeddings for user accounts.
  \item Evaluate the performance of the embedding alignment using metrics such as cosine similarity, Mean Average Precision (MAP) or alignment accuracy.
  \item Implement the chosen embedding algorithm and test it on a dataset of user pairs $(u_{ik}, u_{jl})$ with corresponding embeddings.
  \item Use the embedding model to predict the likelihood of a pair of user accounts representing the same individual for new, unseen data.
\end{itemize} \\ \hline
\end{tabular}
\label{tab:Net-Al-Embeddings}
\end{table*}
In the research described in \cite{qin2020two}, the relationship strength is measured using an improved weighted graph model. First, the authors represent the social network as a weighted graph, where the weight relates to the user interactions. Then, they suggest using the CNIL (Common Neighbors and Internal Links) index, which may also be used to represent social links, to quantify the weight. With data gleaned from second-order neighbors, the CNIL index aims to improve the CN (Common Neighbors) index. The authors then divide these social links into strong ties and weak ties based on the weights by taking into account the social theory that asserts strong ties have a tendency to draw close friends into the same social circles. Nodes that are indirectly connected by two strong linkages are given special consideration.

In \cite{trung2020adaptive} authors suggest GAlign as a method for unsupervised network alignment that does not require any prior understanding of the relationships between the networks (aka anchor links). Given that this paradigm is based on rich network data and multi-dimensional embeddings. Regardless of its modality, information is tied to the network structure and to node properties that can be expressed, such as age, email address, and marital status.
The model proposed in \cite{xie2018unsupervised} is made to handle diverse social connections, content types, and profile aspects of various OSNs. The fundamental tenet is that each element of a user identification describes the actual identity owner, setting that person apart from other users. The experiment results show that Factoid Embedding outperforms the state-of-the-art methods even without training data.

Another hybrid approach (namely \textit{INFUNE}) is presented in \cite{chen2020novel}. The information fusion component and the neighborhood enhancement component make up the model. A weighted sum of node similarity and neighborhood similarity is evaluated as the unified similarity for user identification linkage. The raw properties of users, including structure, profile, and content, coupled with known anchor linkages, are initially pre-processed as distinct similarity matrices. A set of encoders and decoders are used by the information fusion component to combine heterogeneous data and produce discriminative node embeddings for preliminary matching.
%%NUOVA-NUOVA
\subsubsection{Semi-supervised and supervised methods}
Some authors proposed a semi-supervised approach based on the trustworthiness of certain users. Authors in \cite{li2018user} examine the influence of a user's social network. The main idea is that if the majority of someone's closest friends believe certain accounts across different networks belong to them, those accounts are presumed to be theirs. An Authority-Trustworthiness Analysis Model has been developed to determine each friend's authority and the reliability of their verdict. The authors address the UIL problem only if structural information (links between users) is available.

Recently, there has been increasing interest in applying graph embedding techniques for semi-supervised learning. These methods enable the extraction and representation of the structural properties of vertices in networks through low-dimensional latent vectors \cite{goyal2018graph}. Certain approaches are designed to position vectors closer together in the latent space if the corresponding vertices exhibit greater similarity in their identities during the vector representation process.

In 2022 scholars in \cite{yang2022anchor} suggest a multiple consistency-based anchor link prediction approach (MC). It employs intralayer structure information through network representation learning and interlayer structure information in an iterative manner. A matrix factorization-based network representation learning technique is used to learn embedding vectors that include global structural properties of nodes when employing the intralayer structural information. The mapping function of a radial basis neural network is then trained to map embedding vectors from many spaces to a single space. Finally, by taking into account both the interlayer and intralayer structures, the anchor linkages between node pairs are predicted.

Using common profile attributes, authors in \cite{mu2016user} employ a latent user space to solve the UIL problem starting from basic profile attributes such as gender, nationality, birthday, marital status, degree, work experience, location, and educational background. Each real person has a corresponding point into the latent user space that they relate to. If a real user maintains accounts on several social media sites, each one is just seen as a projection of the real person underneath. More specifically, anything that can be seen about a real person on a social network, such as their profile traits, is a projection of that person that is bound by the feature structures that the platform offers.
It follows from this model that when the data from different platforms are projected to this latent space, the data points of the same user should be close to each other (ideally, they should be projected to a single data point). In essence, the more different the two users, the greater the distance between their data points in the latent user space

\section{UIL as a Classification Problem}
\label{Section:Class}
This section will explore and analyze modern methods that treat UIL as a binary classification problem where the goal is to predict whether two nodes (one from each network) represent the same user. Labeled data (pre-aligned user pairs - SAPs) are used to train classifiers like logistic regression, support vector machines (SVM), or deep learning models. Strategies proposed in the literature were mainly \textit{data-oriented}. Consequently, we will summarize the main strategies associated with single data categories, such as  \textit{social connections data, profile, and content data, behavioural data, spatio-temporal data}, and \textit{network traffic data} as well as strategies that apply to combinations of these data categories. 
As known, raw data provides the foundational elements extracted from social media, while features are processed, transformed, and utilized for analytical purposes that are task-specific.

\subsection{Features-based strategies}
Table \ref{tab:Class-features} illustrates an overview of how this process operates with a given set of input data. It begins with feature extraction, where relevant attributes such as user profiles, behaviors, and content are gathered from various networks. Following this, classification algorithms like decision trees or SVMs are employed to classify user accounts, aiming to identify which accounts belong to the same individual based on these features. Evaluation of these methods typically involves assessing classification accuracy and other performance metrics, such as precision, recall, and F1-score, to gauge the effectiveness of the User Identity Linkage.
%sottotask_1
\begin{table*}[t!]
\caption{UIL as a Classification problem Features-based}
\centering
\begin{tabular}{|p{13cm}|}
\hline
\textbf{UIL as a Classification problem Features-based} \\ \hline
\textbf{GIVEN:} \\
\begin{itemize}
  \item Two or more social network platforms denoted as $P_1, P_2, \ldots, P_n$.
  \item Each platform $P_i$ contains a set of user accounts denoted as $U_i = \{u_{i1}, u_{i2}, \ldots, u_{im_i}\}$.
  \item Each user $u_{ij}$ has associated attributes or features, such as:
  \begin{itemize}
    \item Username, email address, or other identifiers
    \item Profile information (e.g., name, age, location)
    \item Behavioral patterns (e.g., posting frequency, types of interactions)
    \item Metadata (e.g., interests, groups, friends)
  \end{itemize}
\end{itemize}

\textbf{TASK:} \\
\textbf{Problem Definition:} \\
\begin{itemize}
  \item Define the task as a binary classification problem, where the goal is to predict whether a pair of user accounts from different platforms represent the same individual or not.
\end{itemize}

\textbf{Formalization:} \\
\begin{itemize}
  \item Let $U_i$ and $U_j$ be the sets of user accounts on platforms $P_i$ and $P_j$ respectively.
  \item For a given pair of user accounts $u_{ik} \in U_i$ and $u_{jl} \in U_j$, define the task as predicting the binary label $y_{ik,jl}$ where:
  \[
  y_{ik,jl} = 
  \begin{cases} 
  1 & \text{if } u_{ik} \text{ and } u_{jl} \text{ are the same individual} \\ 
  0 & \text{if } u_{ik} \text{ and } u_{jl} \text{ are different individuals}
  \end{cases}
  \]
  \item The task can be denoted as learning a function $f : (u_{ik}, u_{jl}) \to y_{ik,jl}$.
\end{itemize}

\textbf{Feature Representation:} \\
\begin{itemize}
  \item Define a set of features $X_{ik,jl}$ representing the similarity or dissimilarity between the attributes of user accounts $u_{ik}$ and $u_{jl}$.
  \item Features can include:
  \begin{itemize}
    \item Direct matching features (e.g., identical identifiers)
    \item Behavioral features (e.g., posting frequency, interactions)
    \item Attribute similarity features (e.g., name similarity, location proximity)
    \item Graph-based features (e.g., shared friends, groups)
    \item Textual features (e.g., similarity of profile descriptions, posts)
  \end{itemize}
\end{itemize}

\textbf{Model Selection and Evaluation:} \\
\begin{itemize}
  \item Formulate the learning objective as optimizing a binary classification model.
  \item Choose an appropriate model for binary classification, such as logistic regression, SVM, or neural networks.
  \item Evaluate the performance of the model using metrics such as accuracy, precision, recall, F1-score, or Receiver Operating Characteristic (ROC) curve.
  \item Implement the chosen model and train it on a labeled dataset of user pairs $(u_{ik}, u_{jl})$ with corresponding labels $y_{ik,jl}$.
  \item Use the trained model to predict the likelihood of a pair of user accounts representing the same individual for new, unseen data.
\end{itemize} \\ \hline
\end{tabular}
\label{tab:Class-features}
\end{table*}
\subsubsection{Profile attributes and contents data}
One of the very first and simple approaches is presented in \cite{perito2011unique}. This study investigates the feasibility of connecting user profiles solely based on their usernames. It makes sense that the "entropy" of the username string itself has a significant impact on the likelihood that two usernames correspond to the same actual individual. This research work, which is based on crawls of actual web services, demonstrates that a sizable part of user profiles can be connected using usernames. In a more recent approach \cite{yuan2021user}, authors utilize Back Propagation (BP) to change the issue into a mapping problem across several social networks, which reduces the distance between username feature vectors and, to a certain extent, eliminates the need for marked user pairs and training iterations. According to the authors, 59\% of users share the same username across several social networks, and such information is typically readily available.

Employing different attributes than usernames, authors in \cite{wang2020user} examine the user profile connection across various social platforms by putting forth an effective and efficient model named MCULK, which is different from the prior work. The model has two essential parts: 1) Producing a similarity graph using profile attributes - like username, bio, etc. - that match user profiles. Then authors utilize locality-sensitive hashing (LSH) to block user profiles and only measure the similarity for those inside the same bucket in order to accelerate the creation. 2) Connecting user profiles using a network of similarity. 
 
Extending the use of common profile attributes to images, in the \textit{Hiding Your Face Is Not Enough (HYFINE)} model, a User Identity Linking model that fully utilizes photos in profiles, is presented in \cite{ranaldi2020hiding}. The HYFINE model is divided into two sections: (1) The corpus extraction method; and (2) The classification system HYFINE-c, which fully utilizes pictures together with other features to categorize two profiles to determine if these profiles are two different identities of the same user. HYFINE-e offers the option to select several profile features for each retained profile. First name/last name, free text about the person, gender, location, profile image, and the last five posts with likes, comments, shares, and, if relevant, retweets were retrieved by the authors.

Other approaches focused on the user content. In the study by \cite{benkhedda2020identity}, authors make an effort to develop a comprehensive framework for user identity linkage across various social networks that is based solely on easily accessible textual user-generated content. Employing deep learning for NLP (Natural Language Processing), authors in \cite{sha2016matching} change the challenge to a straightforward document categorization problem utilizing text content that people have posted on a social network. Authors construct a word vector space first using the messages that all users have posted. Then, they create a document vector space. To create a user's word vector, they use Word2vec. Additionally, there are two ways to create a document vector: 1) Mean-pooling: add the word vectors in users' messages to obtain the average value, which is then used as the document vector; 2) doc2vec. Similarly, in \cite{li2019practical} authors collect all the information of a user page including username, user profile, user content, and user behavior. In this step, they use data preprocessing such as removing inactive users, "zombie" users, to reduce disturbance. Then the authors turn the named entity extracted from user information into 10 categories: Location, Name, Band, Company, Facility, Product, Sport, URL, Date, and others. All useful attributes of a profile can be classified into one category above identified. For example, user address and workplace are considered as Spatio-Temporal data, while keeping e-mail address and personal website in the URL category. Then the authors allocate weight to different entities for distinguishability: (1) user-generated- content: original tweets are more believable than forwarded ones; (2) Part of one site: profile is compared to user- generated-content; (3) Different social networks: information in LinkedIn are generally more serious than in Facebook. Here authors compare the similarity between profiles across different platforms. Then the authors transform the question into a two-class classification task. 

Finally, for criminal search purposes, authors in \cite{zhou2019translink} created a targeted identity resolution method that uses a dataset and a single name to search for false identities of a certain target user. Data on the person's first and last names, gender, date of birth, ethnicity, and both the person's home address and the scene of the crime are necessary for the methodology.

\subsubsection{Spatio-Temporal data}

Recently, cross-device and cross-domain UIL have received a lot of attention. Getting user linkage with spatio-temporal data produced by the numerous GPS-enabled gadgets is a key area of research. The spatial-temporal localization of user actions is used in \cite{feng2019dplink} to explore a more general method of linking user IDs. The essential insight is that authors can connect any online services a user uses to their physical presence, which is determined by time and location. In \cite{chen2017exploiting} authors suggest a brand-new STUL (Spatio-Temporal User Linkage) model to address the issue, which has the following two elements: 1) Using a density-based clustering method to extract the spatial features of users and the Gaussian Mixture Model to recover the users' temporal features. Then the authors give the retrieved features varied weights by downplaying the similar features and emphasizing the discriminative features in order to link user pairs more precisely; 2) Suggesting methods for comparing users based on the attributes retrieved, and then return the pair-wise users with similarity scores that are higher than a specified threshold. Also, the authors in \cite{ding2020user} observe that users with similar mobility patterns frequently check in at a few common sites. So their algorithm (CP-Link) entails two phases: (1) Stay Region Building: Using a DP-based clustering technique, they first create unique stay zones for every user in order to extract their movement patterns; (2) They perform UIL based on IDWT (Inverse Discrete Wavelet Transform). To compare the stay areas of cross-domain users, we offer the IDTW time series similarity matching model. After finishing UIL, the user pair with the highest similarity is chosen as the output-linked pair. However, this class of approaches usually suffers limitations due to the complexity and dimension of location and time series representations and management. Addressing these issues, the work in \cite{chen2018effective} presents a general approach that takes into account both efficacy and efficiency at the same time while performing user account linking with location data. The authors create an innovative approach based on kernel density estimation to address the data sparsity issue. They divided an area into grid cells and focused on each cell to address the issue of data missing. In addition, the author developed an entropy-based weighting mechanism for the grid cells to address the problems brought on by negative coincidence.

To summarize, all the discussed methods focus on linking user identities across various devices or domains by utilizing spatio-temporal data. They extract features or patterns from location data to establish connections between user IDs. A common step in these methods is the comparison of users to identify similarities or matches.

The main differences rely on:
\begin{itemize}
    \item how they handle location data (grid cells, stay zones, spatial features).
    \item how they compare users (similarity scores, IDTW time series similarity matching, entropy-based weighting).
     \item the techniques they use for extracting features or patterns (density-based clustering, Gaussian Mixture Model, DP-based clustering, kernel density estimation).
    \item the implementation trough single/multi phases, the STUL model and CP-Link algorithm involve multiple steps, while the spatial-temporal localization method and kernel density estimation approach do not have explicit phases.
    \item how they address data sparsity and missing data issues, the kernel density estimation approach explicitly addresses both, while the other methods do not mention these issues.

\end{itemize}
\subsubsection{Network traffic data}

According to the author \cite{liu2020user}, each user action generates one or more packets from network traffic data, and these packets of cookies and other information carry a significant amount of user account correlation. In a short period, the user's actions across several network service platforms will be reflected in the network traffic data in this way, causing previously unconnected data to exhibit a particular association. As a result, the network traffic data contains much more useful hidden association information than the material conveyed by the Web. In the case of a dynamic variable IP address, this method may reliably correlate numerous accounts of a user in network traffic with more than 85\% accuracy using only the IP address and the online time. Including also IP-based features, temporal features, geo-based features, device-based features, and household similarities (information of people in one household or organization), an identity graph is built \cite{jalali2018learning} by discovering identity relationships using both online data traffic and offline data logs to establish links between different identities allowing for richer insights into the consumer. Then authors use a machine learning-based approach for the Identity Graph to address the UIL task.
\subsubsection{Mixed data}
According to authors in the work \cite{zhang2016social}, traits derived from the nickname have been frequently employed to identify social connections on various social media platforms. Few works, according to the authors, have relied solely on moniker traits for identification. The authors then take into account hometown similarities while noting that various social networks may have access to various forms of location data. They discuss how to calculate hometown/location similarity using various forms of location data. Finally, the authors take into account user-friendliness in order to identify the same user across several social networks. Expanding this approach, authors in \cite{nurgaliev2020matching} propose an algorithm based on network topology and just the full-name feature of the nodes. The authors expect that the user profile contains at least the full-name of a user. They formulate the problem of linking user accounts from
two social networks with limited profile data as an instance of maximum subgraph matching with the noisy name feature, i.e., full name of a user.

The authors of \cite{bennacer2014matching} presented an algorithm that iteratively matches profiles across social networks based on people who publish the linkages to their numerous profiles using the network structure and publicly available personal information.

Building on prior research, in \cite{quercini2017liaison} authors present LIAISON (reconciLIAtion of Individual profiles across Social Networks), an algorithm that iteratively reconciles profiles across \textit{n} social networks based on the presence of people who disclose the links to their various profiles. LIAISON uses the network topology and publicly available personal information. 

Authors in \cite{shao2021locate} formalize the association between geo-locations and texts instead of using similarity evaluation, and they suggest a brand-new User Identity Linkage framework for locating users across networks. Moreover, by using external text-location pairs, the model can solve the label scarcity issue.

A variety of features are used by authors in \cite{chatzakou2020user} to conduct UIL within the same social network. Two datasets are used by the authors. The first one discussed abusive behavior on Twitter, and the second one was about terrorism. The authors took into account a number of features, including: a) Profile features extrapolated from a user's profile, such as demographic data, a biography, and an avatar; b) Activity features pertaining to a user's posting behavior, such as the number of posts, replies, and mentions; c) Linguistic features extrapolated from users' posted content that may be used to model users with respect to writing style or topics of interest; d) Network properties derived from interactions in social networks between users. 

\subsection{embedding-based strategies}
Table \ref{tab:Class-embeddings} provides a general scheme detailing how this type of task works given a set of input data. It begins with embedding generation, where network data is transformed into dense vector representations that capture both local and global structural information. Following this, classification algorithms like neural networks or support vector machines are employed to classify user accounts, aiming to identify which accounts belong to the same individual based on these embeddings. The effectiveness of these methods is generally measured by examining classification accuracy alongside key performance indicators like precision, recall, and the F1-score.
%sottotask_2
\begin{table*}[t!]
\caption{UIL as a Classification problem Embedding-based}
\centering
\begin{tabular}{|p{13cm}|}
\hline
\textbf{UIL as a Classification problem Embedding-based} \\ \hline
\textbf{GIVEN:} \\
\begin{itemize}
  \item Two or more social network platforms denoted as $P_1, P_2, \ldots, P_n$.
  \item Each platform $P_i$ contains a set of user accounts denoted as $U_i = \{u_{i1}, u_{i2}, \ldots, u_{im_i}\}$.
  \item Each user $u_{ij}$ has associated attributes or features, such as:
  \begin{itemize}
    \item Username, email address, or other identifiers
    \item Profile information (e.g., name, age, location)
    \item Behavioral patterns (e.g., posting frequency, types of interactions)
    \item Metadata (e.g., interests, groups, friends)
  \end{itemize}
\end{itemize}

\textbf{TASK:} \\
\textbf{Problem Definition:} \\
\begin{itemize}
  \item Define the task as a binary classification problem, where the goal is to predict whether a pair of user accounts from different platforms represent the same individual or not.
\end{itemize}

\textbf{Formalization:} \\
\begin{itemize}
  \item Let $U_i$ and $U_j$ be the sets of user accounts on platforms $P_i$ and $P_j$ respectively.
  \item For a given pair of user accounts $u_{ik} \in U_i$ and $u_{jl} \in U_j$, define the task as predicting the binary label $y_{ik,jl}$ where:
  \[
  y_{ik,jl} = 
  \begin{cases} 
  1 & \text{if } u_{ik} \text{ and } u_{jl} \text{ are the same individual} \\ 
  0 & \text{if } u_{ik} \text{ and } u_{jl} \text{ are different individuals}
  \end{cases}
  \]
  \item The task can be denoted as learning a function $f : (u_{ik}, u_{jl}) \to y_{ik,jl}$.
\end{itemize}

\textbf{Embedding Representation:} \\
\begin{itemize}
  \item Utilize graph embedding techniques to represent user accounts $u_{ik}$ and $u_{jl}$ as low-dimensional vectors $e_{ik}$ and $e_{jl}$ respectively.
  \item Embeddings capture structural properties and attribute similarities of user accounts in a latent space.
  \item Techniques such as node2vec, DeepWalk, or graph neural networks (GNNs) can be used to generate embeddings.
\end{itemize}

\textbf{Feature Representation:} \\
\begin{itemize}
  \item Define a set of features $X_{ik,jl}$ based on the embeddings $e_{ik}$ and $e_{jl}$ representing the similarity or dissimilarity between the user accounts.
  \item Features can include:
  \begin{itemize}
    \item Cosine similarity of embeddings
    \item Euclidean distance between embeddings
    \item Dot product of embeddings
    \item Concatenation of embeddings
  \end{itemize}
\end{itemize}

\textbf{Model Selection and Evaluation:} \\
\begin{itemize}
  \item Formulate the learning objective as optimizing a binary classification model.
  \item Choose an appropriate model for binary classification, such as logistic regression, SVM, or neural networks.
  \item Evaluate the performance of the model using metrics such as accuracy, precision, recall, F1-score, or Receiver Operating Characteristic (ROC) curve.
  \item Implement the chosen model and train it on a labeled dataset of user pairs $(u_{ik}, u_{jl})$ with corresponding labels $y_{ik,jl}$.
  \item Use the trained model to predict the likelihood of a pair of user accounts representing the same individual for new, unseen data.
\end{itemize} \\ \hline
\end{tabular}
\label{tab:Class-embeddings}
\end{table*}

\label{sec:core}
 
%As an example, Figure \ref{fig:feat_ext} illustrates the relationship between raw data and features in the specific scenario of UIL task across X (Twitter) and Instagram. 
 %All the state-of-the-art methods will be further cataloged, distinguishing between supervised and unsupervised approaches. 

 \begin{table*}[]
\centering
 %\tiny
%\setstretch{0.5} % Adjust line spacing
 \label{tab:features_works} 
  \caption{Data Categories most used in the literature with the corresponding works listed. }
 \begin{tabular}{lc}

 \hline {\textbf{Data Category}} & 
 {\textbf{References}} \\
 \hline
 \hline 
 {Social connections data}  & \thead{
 \cite{yang2022anchor},\cite{zhou2017structure},\cite{zhang2019graph},
 \cite{guo2020user},\cite{kaushal2020nexlink},\cite{fu2020deep},
 \cite{li2018user}, \\ \cite{zhou2017structure},\cite{man2016predict},\cite{wang2020hyperbolic}, \cite{zhang2014meta},\cite{wang2019community},
 \cite{li2019adversarial},\cite{amara2022cross}}\\

 \hline 
 {Profile attributes and contents data}  & 
 \thead{\cite{perito2011unique}, \cite{yuan2021user},\cite{wang2020user}, \cite{mu2016user},\cite{ranaldi2020hiding},\\ \cite{benkhedda2020identity},
 \cite{sha2016matching},\cite{li2019practical}, \cite{zhou2019translink}}\\
 
\hline 
 {Behavioural data}  & 
 \thead{\cite{liu2015structured}, \cite{sun2019dna},
 \cite{quercini2017liaison}, \cite{shao2021locate}}
 \\
 \hline 

 {Spatio-temporal data}  & 
 \thead{\cite{feng2019dplink}, \cite{chen2017exploiting}, \cite{ding2020user}, \cite{chen2018effective}}\\
 
 \hline
 {Network traffic data}  & 
 \thead{\cite{liu2020user}, \cite{jalali2018learning}}
 \\
  \hline   
 {Mixed data}  & 
 \thead{\cite{zhang2016social}, \cite{nurgaliev2020matching}, \cite{bennacer2014matching}, \cite{chatzakou2020user}, \cite{hu2021semi},
 \cite{mu2016user}, \cite{wang2019user}, \\ \cite{su2018master}, \cite{trung2020adaptive}, \cite{xie2018unsupervised}, \cite{chen2020novel}, \cite{qin2020two}, \cite{wang2019anchor}, \cite{zhang2015multiple}, \\ \cite{lei2018catching}, \cite{zhong2018colink}
}
 \\

 \hline
 %\hline
   \end{tabular}
 %  }
\end{table*}

%\Figure[t!](topskip=0pt, botskip=0pt, midskip=0pt){fig1.png}
%{ \textbf{Magnetization as a function of applied field.
%It is good practice to explain the significance of the figure in the caption.}\label{fig1}}
\subsubsection{Social connections data (Graph-based feature)}
One of the most used information for UIL tasks are Graph-based features (e.g., shared friends, groups). In this case, the structure of a dataset generally consists of a list of ID pairs from an OSN.
Each pair including user $u_a$ and user $u_b$ represents the relationship between the two users. Depending on the OSN considered, such a relationship can be of mutual friendship as in the case of Facebook, or, for instance, the first user in the pair follows the second one. This is the case in which the social relationship between users is not mutual so the social connection is oriented (e.g., followees or followers on Twitter). 

Traditional methods often rely on either interlayer structures, which refer to the connections between nodes across different layers or networks, or intralayer structures, which refer to the connections between nodes within the same layer or network. As a result, they do not fully utilize both interlayer and intralayer structures for anchor link prediction. 

In \cite{zhou2018deeplink} the author proposed a model to capture local and global network structures. DeepLink samples the networks and learns to encode network nodes into a vector representation. This information may then be utilized to align anchor nodes using deep neural networks. The policy gradient approach is used to learn how to transmit knowledge and update the linkage utilizing a dual learning-based paradigm. 
The authors in \cite{zhang2019graph} also include local and global properties of a network. Specifically, the first part of the proposed model encodes the social network's graph architecture into node features. Node embeddings, a common approach in network representation learning, is what this feature learning procedure entails. By projecting the network structure to the low-dimensional node space, this embedding serves to retain both the global and local graph connection patterns, resulting in rebuilt networks that are reasonably similar to the original networks and can be easily compared for UIL predictions. Recently, a neural tensor network-based approach called NUIL \cite{guo2020user} employs the Random Walks and Skip-gram models to incorporate the network structure in a low-dimensional vector space. In NUIL, a neural tensor network model, which is better able to express the relationships between users, takes the role of a conventional neural network model. The model first creates several social sequences for each user in several rounds of random walks, encoding the social ties between users in the social networks, before embedding users into a latent space to compare latent vectors. Assessing the individual contribution of local and global properties on the same social network, authors in \cite{kaushal2020nexlink} propose the NeXLink node embedding framework, which consists of three parts.
The local structure of nodes within the same social network is first preserved in order to produce local node embeddings. The global structure, which is present in the form of the common friendship displayed by nodes involved in CNLs across social networks, is preserved in order to learn the global node embeddings. Thirdly, local and global node embeddings are integrated, to keep local and global structures and make it easier to identify CNLs across social networks. Finally, in \cite{fu2020deep} authors embed graph vertices into low-dimensional vector space to investigate a multi-granular user identity alignment system. First, the higher-order structural qualities, and second, the SAP-oriented structural consistency in the topology of social networks, are preserved by a framework's two granular layers. This framework is what authors refer to as a Multi-granular Graph Embedding framework (MGGE). Furthermore, the authors extended the model—known as the "DeepMGGE" model to include its capacity to capture the non-linear structural characteristics of SAP-oriented structural consistency.

To provide a robust method, the authors of \cite{man2016predict} propose a novel supervised model called PALE that uses network embedding with awareness of observed anchor links as supervised information to capture the intrinsic structural regularities of networks rather than working directly on them as most existing methods, unsupervised or supervised, do. As a drawback, the effectiveness of the method is sensitive to the high dimension and sparsity of networks. Avoiding dimensionality limitations, authors in \cite{wang2020hyperbolic}  discovered that hyperbolic geometry has the advantage of describing network hierarchical structure, whereas Euclidean geometry does not, which is prompted by current developments in geometry representation learning. As a result, the authors first discuss how social networks and hyperbolic space are related in this work. After that, the authors provide a brand-new "HUIL" hyperbolic geometry representation learning model for user identification linking across social networks.

Approaches based on clusters or community similarities have also proved to be effective on common UIL datasets. For instance, employing the proposed Foursquare-Twitter dataset (\cite{zhang2014meta}) in \cite{wang2019community} authors put forth a fresh embedding-based method that takes into account and makes use of both individual and community similarity by concurrently maximizing both in a single loss function.
Authors in \cite{li2019adversarial} accomplish identity alignment at the distribution level and take a holistic perspective of all the identities in a social network. The identities of the same natural person will be clustered together in the proposed model, which transforms the identity distribution in Twitter space by a set of operations (such as transposing) to minimize the distance between it and the identity distribution in Facebook. The authors' transformation of the social network alignment problem to the learning of the operation to minimize the distance between two distributions is motivated by isomorphism. In \cite{amara2022cross}, in contrast to earlier efforts, the suggested model takes into account the multi-network scenario to encapsulate various anchor users' network architectures. For each social network, the authors suggest a high-dimensional base embedding and a low-dimensional social edge embedding to capture the various structural details of an anchor user from various social networks. In particular, using one of three possible aggregator functions—mean, max-pooling, or LSTM—with a self-attention mechanism, authors develop a function that creates social edge embeddings by sampling and averaging structural data from an anchor user's neighborhood inside various social networks. As a downstream task, link prediction is utilized to assess how well the learned embeddings work.

%Summarizing: Recent approaches (MC, DeepLink, the model in \cite{zhang2019graph}, NUIL, NeXLink, MGGE) use network representation learning to embed nodes into a vector space considering both local and global network structures. 
Summarizing, the main differences among these approaches rely on:
\begin{itemize}
            \item how they handle the network structure (interlayer, intralayer, local, global); 
            \item the use of \textbf{supervised learning} (PALE, HUIL, NeXlink, NUIL), \textbf{semi-supervised learning} (DeepLink); 
            \item the techniques they use for embedding (matrix factorization, deep neural networks, radial basis neural network, random walks, skip-gram, hyperbolic geometry).
      \end{itemize}      
Moreover, the strategy in  \cite{li2019adversarial} considers the distribution of identities in a social network, while the model in \cite{amara2022cross} considers a multi-network scenario, which is not the case for all the others.

\subsubsection{Behavioural data}
Authors in \cite{liu2015structured} propose a solution framework, HYDRA, which consists of three steps: (I) modeling heterogeneous behavior by long-term topical distribution analysis and multi-resolution temporal behavior matching against high noise and information missing. The behavior similarity is described by multi-dimensional similarity vector for each user pair; (II) building structure consistency models to maximize the structure and behavior consistency on users’ core social structure across different platforms, thus the task of identity linkage can be performed on groups of users, which is beyond the individual level linkage in the previous study; and (III) proposing a normalized-margin-based linkage function formulation, and learn the linkage function by multi-objective optimization where both supervised pair-wise linkage function learning and structure consistency maximization are conducted towards a unified Pareto optimal solution.

The order of friending in actual dynamic social networks is utilized by the authors in \cite{sun2019dna}. In reality, social psychology research shows that an individual's friendship growth across social networks is predominately deterministic rather than stochastic \cite{Friedkin1998}.

\subsubsection{Mixed Data}

With the aid of a dynamic hypergraph neural network, in \cite{hu2021semi} the feature extraction model learns node embeddings from topology space and feature space. In the WGAN training phase, the network alignment model employs a new sampling technique that places more emphasis on sample-level data. The outcomes of thorough tests conducted on the real-world dataset confirm the efficiency of the suggested framework.

In \cite{mu2016user} employing the ego networks of two users as input, authors formalize the user alignment across social networks as a classification problem. In order to align the users, the authors propose a graph neural network model called MEgo2Vec to describe the matched ego networks of the two users as a low-dimensional real-valued representation. The representation is divided into two parts: one is an embedding from the target user pairs' and their neighbor pairs' attributes, and the other is an embedding from the matching ego network's topologies. In a later work \cite{wang2019user}, the authors model the topics of user interests to represent the content information in different social networks at the same granularity and filter out the noise. Second, they capture friend-based (i.e., structure) and interest-based (i.e., content) user co-occurrence in linked heterogeneous networks using four types of sub-networks (i.e., user-user intra/inter-network and user-topic intra/inter- network). Third, they learn effective user representations by embedding the sub-networks into a unified low-dimensional space. Also in \cite{su2018master} - where the authors propose the MASTER framework - are integrated attribute and structure embedding for reconciliation across several social networks. In this framework, in order to define the problem as a unified optimization, authors first build a novel constrained dual embedding model by simultaneously embedding and reconciling several social networks.

The majority of approaches ignore the social network attribute data. In order to solve the issue, authors in \cite{wang2019anchor} suggest a brand-new semi-supervised network-embedding approach. Each node of the numerous networks is represented in the model by a vector for predicting anchor connections, which is learned with knowledge of the observed anchor links as semi-supervised information and input, as well as topology structure and attributes. The suggested model outperforms several state-of-the-art methods, as shown by experimental findings on real-world data sets.

In \cite{zhang2015multiple} authors study the M-NASA problem to identify the anchor links among multiple anonymized social networks. In addition to its significance, the M-NASA problem is a brand-new problem that is entirely distinct from previous efforts. The suggested procedures are as follows: (1) supervised anchor link inference across social networks, which focuses on inferring the anchor links between two social networks with a supervised learning model; (2) network matching, which investigates various heuristics to match two networks based on the known existence probabilities of potential correspondence relationships; (3) entity resolution, which aims at discovering multiple references to the same entity in one single database with a relational clustering algorithm; (4) cross-media user identification connects users from different networks based on data from multiple node attributes produced by users' social interactions.

The framework proposed in \cite{lei2018catching} uses word2vec \cite{mikolov2013distributed} and DeepWalk \cite{perozzi2014deepwalk} to first turn all textual and structural user data into low-dimensional latent spaces, then it integrates various user features and predicts empty data fields using a late fusion technique and computations based on cosine similarity. The outcomes demonstrated that by enhancing and modernizing data sources as needed, the methodology may successfully capture dynamic user data and improve the performance of identity linkage models.

\begin{table*}[]
\caption{A synthesis of the most innovative algorithms mentioned in this survey.}
\label{tab:algo}
\vspace*{5mm}
\fontsize{10pt}{10pt}\selectfont
 \centering
\begin{tabular}{l|c|l|c|c}
\hline
\multicolumn{1}{c|}{\textbf{Algorithm Name}} &
%\multicolumn{1}{c|}{\textbf{\begin{tabular}[c]{@{}c@{}}Algorithm \\ Name \end{tabular}}} &
%\multicolumn{1}{c|}%{\textbf{Sup/UNSup/SemiSup}}
\begin{tabular}[c]{@{}l@{}}\textbf{Sup/UNSup}\\ \textbf{Semi-Sup}\end{tabular}&
%\multicolumn{1}{c|}{\textbf{\begin{tabular}[c]{@{}c@{}}Algorithm \\ Name \end{tabular}}} &
  \multicolumn{1}{c|}{\textbf{Description}} &
  \textbf{Data Category} &
  \textbf{Ref.} \\ \hline \hline
NUIL &  Sup &
  \begin{tabular}[c]{@{}l@{}}Neural tensor network-\\ based approach to UIL\end{tabular} &
  Social Connection &
  \cite{guo2020user} \\ \hline
NeXLink & Sup &
  \begin{tabular}[c]{@{}l@{}}Node Embedding \\ Framework for Cross-\\ Network Linkages \\ Across Social \\ Networks\end{tabular} &
  Social connection &
  \cite{kaushal2020nexlink} \\ \hline
DeepMGGE & Sup &
  \begin{tabular}[c]{@{}l@{}}Deep multi-granularity graph \\ embedding for user \\ identity linkage \\ across social networks\end{tabular} &
  Social connection &
  \cite{fu2020deep} \\ \hline
FRUI-P & UNSup &
  \begin{tabular}[c]{@{}l@{}}Friend Relationship-based \\ User Identification algorithm\\  without Prior knowledge\end{tabular} &
  Social connection &
  \cite{zhou2017structure} \\ \hline
PALE &  Sup &
  \begin{tabular}[c]{@{}l@{}}Predict Anchor Links \\ across Social Networks \\ via an Embedding\end{tabular} &
  Social connection &
  \cite{man2016predict} \\ \hline
STUL & Sup &
  \begin{tabular}[c]{@{}l@{}}Spatio-Temporal User\\  Linkage\end{tabular} &
  Spatio-Temporal &
  \cite{chen2017exploiting} \\ \hline
CP-Link & Sup &
  \begin{tabular}[c]{@{}l@{}}Check-in Patterns for \\ User Identity Linkage\end{tabular} &
  Spatio-Temporal &
  \cite{ding2020user} \\ \hline
HYFINE & Sup &
  \begin{tabular}[c]{@{}l@{}}Hiding Your Face Is Not Enough.\\ A User Identity Linking model \\ that fully exploits images in profiles.\end{tabular} &
  Profile Att. and C. &
  \cite{ranaldi2020hiding} \\ \hline
MEgo2Vec & UNSup &
  \begin{tabular}[c]{@{}l@{}}A graph neural network model\\ to describe the matched \\ ego networks of the two users\\ as a low-dimensional \\ real-valued representation\end{tabular} &
  Mixed & 
  \cite{mu2016user} \\ \hline
MASTER & Semi-Sup &
  \begin{tabular}[c]{@{}l@{}}Across Multiple social networks, \\ integrate Attribute and \\ STructure Embedding \\ for Reconciliation\end{tabular} &
  Mixed & 
  \cite{su2018master} \\ \hline
GAlign & UNSup &
  \begin{tabular}[c]{@{}l@{}}A fully unsupervised network\\ alignment framework \\ based on a multi-order \\ embedding model.\end{tabular} &
  Mixed & 
  \cite{trung2020adaptive} \\ \hline
INFUNE & Semi-Sup &
  \begin{tabular}[c]{@{}l@{}}Information fusion component\\ and the neighborhood \\ enhancement component\end{tabular} &
  Mixed & 
  \cite{chen2020novel} \\ \hline
LIAISON & Sup &
  \begin{tabular}[c]{@{}l@{}}ReconciLIAtion of \\ individuals profiles \\ across social network\end{tabular} &
  Mixed & 
  \cite{quercini2017liaison} \\ \hline
\end{tabular}
\end{table*}

\section{Evaluation Metrics}
\label{sec:evalu}

The second research question \textbf{RQ2}, concerning current SOTA performance, is in line with the practical goal of this investigation, which aims to serve as a practical guide for potential applications. However, being able to compare the plethora of strategies presented in the literature in terms of performance, requires analyzing and comparing them under the same setting which is beyond the scope of this review. 

Trung et al. \cite{trung2020comparative} undertook an endeavor in this regard, presenting a comprehensive empirical examination of the effectiveness of various network alignment methods. They specifically combine a number of cutting-edge network alignment strategies in a comparable way and assess various settings to gauge the individual properties of these techniques with the ultimate goal of providing a benchmark framework useful to identify the best strategy for each scenario.
The benchmark findings, which were achieved using both real data
and synthetic data, are then thoroughly analyzed. The datasets employed are:
Douban, Flickr-lastfm, Flickr-myspace, fb-tw, fq-tw. Interestingly, for several of the models tested, the authors find that on these real datasets, accuracy is equal to 0.00 confirming that each specific scenario has its most suitable network alignment technique since no single technique consistently outperforms all others. 

For these reasons, in this section, we will first give an overview of the evaluation metrics most used in the User Identity Linkage task in its two main formulations adopted in the present survey. Then we will report the results achieved by different research groups (each one implementing a different UIL strategy) applying these metrics on the same dataset, the Forsquare-Twitter dataset. Reported performance values are those declared by the authors in their published research works both directly through numbers and indirectly by charts. 

\subsection{Evaluation Metrics specific for UIL as a Network Alignment task}

\textbf{Alignment Accuracy}.
Alignment accuracy measures the proportion of correctly identified anchor links (ALs) (true matches) out of the total number of possible anchor links. It is calculated as:

\[
\text{Align. Accuracy} = \frac{\text{Number of correctly predicted ALs}}{\text{Total Number of true ALs}}
\]
\\
\textbf{Mean Average Precision (MAP)}.
Mean Average Precision (MAP) is a metric used to evaluate the accuracy of ranking models, considering both the precision of results at different cutoff levels and their order. The average precision (AP) for a single query or user is given by:

\[
\text{AP} = \frac{1}{m} \sum_{k=1}^{m} P(k) \cdot \text{rel}(k)
\]

where:
\begin{itemize}
    \item \( m \) is the total number of true positives.
    \item \( P(k) \) is the precision at rank \( k \).
    \item \( \text{rel}(k) \) is a binary indicator function that equals 1 if the item at rank \( k \) is relevant and 0 otherwise.
\end{itemize}

The MAP is then the average of these AP values across all queries or users:

\[
\text{MAP} = \frac{1}{Q} \sum_{q=1}^{Q} \text{AP}(q)
\]

where \( Q \) is the total number of queries or users.
\\

\textbf{Normalized Discounted Cumulative Gain (NDCG)}
NDCG evaluates the quality of the ranked list of results, giving higher scores to correct matches appearing higher in the ranked list. The DCG at position \( p \) is calculated as:

\[
\text{DCG}_p = \sum_{i=1}^{p} \frac{2^{\text{rel}(i)} - 1}{\log_2(i + 1)}
\]

where \( \text{rel}(i) \) is the relevance score at rank \( i \).

NDCG is the normalized version of DCG, where DCG is divided by the ideal DCG (IDCG), which is the DCG for the ideal ordering of results:

\[
\text{NDCG}_p = \frac{\text{DCG}_p}{\text{IDCG}_p}
\]

IDCG is calculated as:

\[
\text{IDCG}_p = \sum_{i=1}^{|\text{REL}_p|} \frac{2^{\text{rel}(i)} - 1}{\log_2(i + 1)}
\]

where \( |\text{REL}_p| \) is the set of relevant items up to position \( p \).
\subsubsection{Structural Preservation}.
Structural preservation is an evaluation metric used to measure how well the alignment of users across different social networks preserves the structural properties of the original networks. It assesses whether the inherent relationships and connections within the networks are maintained after the linkage.

To evaluate Structural Preservation, we typically look at the consistency of the structural properties, such as the degree distribution, clustering coefficient, and shortest path length, between the original and the aligned networks.

\textbf{Degree Distribution Preservation}.
The degree of a node in a network is the number of connections (edges) it has to other nodes. Degree distribution preservation ensures that the degree of nodes in the aligned network is similar to their degree in the original networks.

\[
D(v) = \text{Degree of node } v
\]

For an aligned node \( v \) across two networks \( G_1 \) and \( G_2 \), the Degree Preservation (DP) can be measured as:

\[
\text{DP} = \frac{1}{|V|} \sum_{v \in V} \left| D_{G_1}(v) - D_{G_2}(v') \right|
\]

where \( D_{G_1}(v) \) and \( D_{G_2}(v') \) are the degrees of node \( v \) and its aligned counterpart \( v' \) in networks \( G_1 \) and \( G_2 \), respectively.

\textbf{Clustering Coefficient Preservation - CCP}.
The clustering coefficient of a node measures the extent to which its neighbors form a complete graph (i.e., are interconnected). Preservation of clustering coefficients ensures that the local neighborhood structure around each node is maintained.

\[
C(v) = \frac{2 \times \text{Number of closed triplets}}{\text{Number of connected triplets}}
\]

For an aligned node \( v \):

\[
\text{CCP} = \frac{1}{|V|} \sum_{v \in V} \left| C_{G_1}(v) - C_{G_2}(v') \right|
\]

\textbf{Shortest Path Length Preservation - SPLP}.
The shortest path length between two nodes is the minimum number of edges required to connect them. Preservation of shortest path lengths ensures that the overall connectivity and distances between nodes are maintained.

\[
\text{SPL}(u, v) = \text{N. of edges in the SP between nodes } u \text{ and } v
\]

For an aligned pair of nodes \( u \) and \( v \) in networks \( G_1 \) and \( G_2 \):

\[
\text{SPLP} = \frac{1}{|E|} \sum_{(u,v) \in E} \left| \text{SPL}_{G_1}(u, v) - \text{SPL}_{G_2}(u', v') \right|
\]

where \( E \) is the set of edges, and \( u' \) and \( v' \) are the aligned counterparts of \( u \) and \( v \) in the other network.

\textbf{Aggregate Structural Preservation Score - Aggregate SPS}
An aggregate score for structural preservation can be calculated by combining the individual preservation metrics, typically using a weighted sum or average:

\[
\text{SPS} = w_1 \times \text{DP} + w_2 \times \text{CCP} + w_3 \times \text{SPLP}
\]

where \( w_1 \), \( w_2 \), and \( w_3 \) are weights that can be adjusted based on the importance of each structural property in the specific application.

\subsection{Evaluation Metrics for both UIL problem formulations}

\begin{table*}[]
 %\small
  \caption{\label{tab:sota_results} SOTA results on the Foursquare-Twitter dataset \cite{zhang2014transferring}. For each metric or variant of a metric (e.g. P@k) we only report the best result provided by the authors. Some results are reported using $\sim$ as they are extracted from graphs published in the works cited and not from explicit numerical reports.}
\vspace*{5mm}
 \centering
 \begin{tabular}{lcccc}
 \hline {Paper} & 
 {Precision} & 
 {Recall}  & 
 {F1}&
 {AUC} \\
 \hline
Ma et al., 2021\cite{ma2022cp}
& 0.63
& 0.60 
& 0.615 
& 0.82   
\\
 
Wang et al., 2018 \cite{wang2018you}
& $\sim$0.8
& $\sim$0.4
& -
& $\sim$0.5   \\

Riederer et al., 2016 \cite{riederer2016linking}
& $\sim$0.85
& $\sim$0.3
& -
& $\sim$0.4   \\

Zhou et al., 2018 \cite{zhou2018deeplink}
& 0.7048
& 0.7914
& -
& 0.991   \\

Zhou et al., 2019 \cite{zhou2019translink}
& 0.7653
& 0.9031
& -
& - \\

Chen et al., 2018 \cite{chen2018effective}
& $\sim$0.4
& $\sim$0.4
& $\sim$0.4
& - \\

Ding et al., 2020 \cite{ding2020user}
& $\sim$0.7
& $\sim$0.6
& $\sim$0.65
& $\sim$0.82 \\

Feng et al., 2020 \cite{feng2019dplink}
& $\sim$0.4 \footnote{In this case only hit precision is reported. }
& -
& -
& - \\

Chen et al., 2017 \cite{chen2017exploiting}
& $\sim$0.8 
& $\sim$0.52
& $\sim$0.62
& - \\

Shao et al., 2021 \cite{shao2021locate}
& -
& -
& 0.8926
& 0.9327 \\
  \hline
  \hline
   \end{tabular}
\end{table*}

%\subsection{Evaluation Metrics}

The following are the common metrics used in literature for different formulations of UIL-related tasks. The metrics are often adapted from study to study with different meanings. Here we collect the generic definition of each metric with a specific comment on the suitability for UIL tasks.

In the field of machine learning for classification tasks, a \textit{True Positive} (TP) is an actual positive sample correctly predicted by a model as positive. Similarly, a \textit{True Negative} (TN) is an actual negative sample correctly predicted as negative. A \textit{False Positive} (FP) is an actual negative sample misclassified as positive. Finally a, a \textit{False Negative} (FN) is an actual positive sample misclassified as negative. Specifically for UIL a TP usually represents a correctly predicted link between users that are actually the same real person. TN a non-existent link correctly non predicted, FP and FN a wrongly predicted link and a non-predicted (but actually existent) link respectively.

\textbf{Accuracy}. Accuracy is the ratio of correct predictions on the total observations and is given by the Equation \ref{eq:accuracy}. Accuracy is one way to measure what percentage of predictions are right.

\begin{equation} \label{eq:accuracy}
Accuracy = \frac{TP+TN}{TP+TN+FP+FN}
\end{equation}

In the specific field of UIL, given a predicted list of pairs with possible links between users from different social networks, accuracy can measure how many links were correctly predicted by the system. However, the TNs component of the equation does not generally contribute (i.e., the interest is in linking two users that are actually the same real person, instead of predicting non-existent links) significantly for UIL-related tasks. 

\textbf{Error rate}. Closely related to Accuracy is the \textit{Error rate}. The definition is given by the Equation \ref{eq:error_rate}. The error rate expresses the percentage of predictions that are wrong.

\begin{equation} 
\label{eq:error_rate}
\resizebox{.85\hsize}{!}{$Error Rate = 1-Accuracy = \frac{FP+FN}{TP+TN+FP+FN}$}
\end{equation}

Depending on how genuine positives and negatives are defined in a multilabel scenario, the definition of this metric may differ. A prediction is deemed accurate (referred to as "subset accuracy") when the projected labels exactly match the actual labels.
Alternately, before the accuracy calculation, predictions can be flattened and condensed to a single-label task. As for the Accuracy, ErrorRate is not often used for UIL-tasks but is more common for similar tasks such as link predictions and friend recommendation.

\textbf{Precision}. Equation \ref{eq:precision} defines \textit{precision} or \textit{sensitivity} as the ratio of true positive (TP) observations to all-around positive predicted values (TP+FP). Precision is the proportion of correctly predicted events among all positively predicted events.

\begin{equation} \label{eq:precision}
Precision = \frac{TP}{TP+FP}
\end{equation}

Precision is one of the most common metric used in UIL-related tasks. When a model provides a list of predicted links between two users from different social networks, this metric measures how many of the predicted links are actually linking the same real person.

\textbf{Recall}. Equation \ref{eq:recall} gives \textit{recall} or \textit{specificity} as the ratio of true positive (TP) observations to all-around positive predicted values (TP+FN). Recall is the ratio of right predictions made overall positive predictions that should have been made.

\begin{equation} \label{eq:recall}
Recall = \frac{TP}{TP+FN}
\end{equation}

For scenarios involving multi-class classification, it is possible to compute the precision and recall for each class label. 
Also, Recall is one of the most common metrics used in UIL-related tasks. In this case, given all the actual links between different users from different social networks, this metric allows measuring how many links were correctly predicted.

\textbf{F1 score}. Equation \ref{eq:f1} illustrates the F1 score, which is the harmonic mean of recall and precision. The maximum precision and recall value of an F1 score is 1, while the lowest value is 0.

\begin{equation} \label{eq:f1}
F1 = 2 \times \frac{Recall \times Precision}{Recall + Precision}
\end{equation}

F1 score is often used in UIL-related tasks to provide, with a single scalar, the performance of the model in predicting links considering both the Precision and the Recall already discussed. 

 \textbf{Matthews Correlation Coefficient (MCC)}. The effectiveness of binary classification techniques is also measured by the Matthews Correlation Coefficient (MCC)\cite{matthews1975comparison}, which collects all the data in a confusion matrix. MCC can be used to address issues with unequal class sizes and is still regarded as a balanced approach. The MCC scales from -1 to 1. (i.e., the classification is always wrong and always true, respectively). Equation \ref{eq:mcc} provides the formula for MCC.

\begin{equation} 
\label{eq:mcc}
\resizebox{.85\hsize}{!}{$MCC = \frac{TP \times TN - FP \times FN}{\sqrt{(TP + FP) (TP+FN)(TN+FP)(TN+FN)}}$}
\end{equation}

For the same consideration about accuracy (i.e., the interest in predicting non-existent links between users from different persons, the case of TNs) MCC is not frequently used in the literature for UIL.

\textbf{AUC}. The region under the ROC curve, which contrasts the true positive rate (TPR) and false positive rate on the ROC curve (FPR). The likelihood that the randomly chosen positive examples rank higher than the negative cases is how the AUC, which we use to assess the validity of similarity rankings, is determined. The returned user account pairs are said to be "positive" in this case if they belong to the same user.

Finally, some specific metrics related to multilabel tasks are Micro and Macro-F1 \cite{manning2008introduction}, and Precision@k and Normalized Discounted Cumulated Gains \cite{liu2017deep}.

\section{Datasets} 
\label{sec:dataset}
To evaluate the performance of the AI methods, agreed benchmark datasets are needed. Until 2016 there were many datasets related to a single social network, but very few datasets were available to be used as ground truth for performing UIL tasks across social networks, thus model validation was challenging. Today the problem of obtaining a comprehensive dataset with different feature spaces still exists but a few steps forward were taken.
Considering that a detailed catalog of all the datasets proposed in the literature misses, we collect in Table\ref{table:dataset} the datasets used in two or more works and built in a previous study. It is worth mentioning that not all of these studies utilized identical partitions of the same dataset. Furthermore, several different metrics are employed, making difficult an objective evaluation of models on the same dataset as clarified above.
In Table\ref{table:dataset} we report when each dataset was first presented along with the year, the reference paper, and the papers reporting the studies where they were used. We also provide the dimension of each dataset (\textit{Size} column) highlighting that \textit{Size} refers to the number of distinct identities linked in the set of social platforms (which can include 2, 3, or more platforms) considered for each dataset. When the cross-dimension of the set is not specified by the dataset authors, we report here the greater number of identities in the set.
It is worth noticing that while 16 datasets are reported in the table, several others - built-up ad-hoc for a single and specific work - are available in the literature. In some cases, also subsets of the datasets shown in the table have been used.
In fact, the large majority of the works on UIL are based on novel datasets presented along with the new proposed approach discussed in the corresponding paper. From Table \ref{table:dataset} it can be seen that the two top-referenced and used datasets in the literature are FT2 and TF2. Here we briefly introduce both of them.

For the \textit{Foursquare-Twitter (FT2)} dataset, the scholars crawled user profiles together with their online tips. Tips and profiles are 94,187 and 5,392 respectively. The total number of places crawled from Foursquare is 38,921 and all tips can be attached to location check-ins. At Foursquare, the two unidirectional follow links that were created from the bidirectional buddy link have replaced the original follow links. Similar to this, Twitter is crawled for 5,223 people and all of their public tweets. The number of tweets crawled by authors is 9,490,707, among which 615,515 tweets contain location check-ins and they constitute around 6.5\% of all the tweets. 297,182 locations in total were gleaned from the tweets.

The second top-referenced dataset is crawled from Twitter and Foursquare. 
The website for Foursquare, a typical location-based social network, was the first one crawled (LBSN). By performing a breadth-first search over the social graph, the authors gathered a dataset of 500 people and 7,504 tips from these users. The latitude and longitude of each tip, as well as the timestamp, are available. The Foursquare network additionally offers information on who a user is following or friending. These connections can show how socially connected the users are.
Then the scholars gathered 500 people, matching the 500 Foursquare users, and 741,529 tweets from the individuals. In the Twitter network, all tweets contain a time stamp, and some tweets also contain a location stamp. In the end, the authors had 34,413 tweets in total with location information (latitude and longitude), which represents 4.6\% of all the tweets we gathered.

\begin{table*}[]
\centering
\caption{List of Datasets for UIL with evidence of year of creation, reference of the research work where it first appeared (Collected in), references to subsequent works in which the dataset was used (Used in), and number of identities (Size).}
\vspace*{5mm}
%\begin{minipage}{\textwidth} \centering
\fontsize{11pt}{11pt}\selectfont
\begin{tabular}{l|l|l|l|l}

\hline
\textbf{Dataset} &
  \textbf{Year} &
  \textbf{Collected in} &
  \textbf{Used in} &
   \textbf{Size} 
  \\ \hline
\begin{tabular}[c]{@{}l@{}}Data Mining -\\ Machine Learning (DM1)\end{tabular} &
  2020 &
  \cite{fu2020deep} &
  \cite{yang2022anchor} &
  22,542
  \\ \hline
\begin{tabular}[c]{@{}l@{}}Douban Online - \\ Douban offline (DD1)\end{tabular} &
  2016 &
  \cite{zhang2016final} &
  \cite{trung2020adaptive} &
  3,906
 \\ \hline
\begin{tabular}[c]{@{}l@{}}Douban Online - \\ Douban offline (DD2)\end{tabular} &
  2012 &
  \cite{zhong2012comsoc} &
  \cite{wang2019anchor} &
  $\sim50,000$
\\ \hline
Facebook (FB1) &
  2009 &
  \cite{viswanath2009evolution} &
  \begin{tabular}[c]{@{}l@{}}\cite{man2016predict},\\ \cite{kaushal2020nexlink}\end{tabular} &
  90,269
\\ \hline
Facebook (FB2) &
  2014 &
  \cite{leskovec2014snap} &
  \cite{li2018user} &
  4,039
\\ \hline
Flickr - LastFM (FL1) &
  2015 &
  \cite{zhang2015cosnet} &
  \cite{wang2019anchor} &
  215,495
\\ \hline
Flickr - MySpace (FM1) &
  2016 &
  \cite{zhang2016final} &
  \cite{trung2020adaptive} &
  10,733
 \\ \hline
Flickr - Twitter (FT1) &
  2013 &
  \cite{yan2013friend} &
  \cite{wang2019user} &
  1,457
 \\ \hline
Foursquare - Twitter (FT2) &
  2014 &
  \cite{zhang2014transferring} &
  \begin{tabular}[c]{@{}l@{}}\cite{ma2022cp}, \cite{wang2018you}, \cite{riederer2016linking},\\
  \cite{zhou2018deeplink}, \cite{zhou2019translink},
  \cite{chen2018effective},\\
  \cite{ding2020user},
  \cite{feng2019dplink},
  \cite{zhang2017link}, \\\cite{chen2017exploiting},
  \cite{shao2021locate}\end{tabular} &
  7,227
  \\ \hline
Instagram - Twitter (IT1) &
  2016 &
  \cite{riederer2016linking} &
  \cite{chen2017exploiting} &
  1,717
  \\ \hline
\begin{tabular}[c]{@{}l@{}}Instagram - Twitter - \\ Google+ (ITG1)\end{tabular} &
  2018 &
  \cite{sharma2018linksocial} &
  \cite{wang2020user} &
  7,729
 \\ \hline
\begin{tabular}[c]{@{}l@{}}Lastfm-MySpace \\ and Livejournal-\\ MySpace (LL1)\end{tabular} &
  2014 &
  \cite{zhang2015cosnet} &
  \cite{zhou2018deeplink} &
  854,498 - 3,017,286
 \\ \hline
%\begin{tabular}[c]{@{}l@{}}Sina Microblog - \\ RenRen (SR1)\end{tabular} &
 % 2015 &
  %\cite{zhou2015cross} &
  %\cite{zhou2017structure} &
  %&
  %NA \\ \hline
Twitter - Flickr (TF1) &
  2017 &
  \cite{sun2017mapping} &
  \cite{lei2018catching} &
  7,109
  \\ \hline
Twitter - Foursquare (TF2) &
  2014 &
  \cite{kong2013inferring} &
  \begin{tabular}[c]{@{}l@{}}\cite{liu2016aligning}, \cite{wang2019community}, \cite{zhang2015integrated}, \\ \cite{zhang2014meta}, \cite{fu2020deep}, \cite{su2018master}, \\ \cite{kaushal2020nexlink},\cite{guo2020user} \end{tabular} &
  500
 \\ \hline
\begin{tabular}[c]{@{}l@{}}Twitter - \\ LiveJournal - YouTube - \\
Flickr\end{tabular} &
  2012 &
  \cite{buccafurri2012discovering} &
  \cite{bennacer2014matching}, \cite{quercini2017liaison} &
  93,169
  \\ \hline
\end{tabular}

%\end{minipage}

\label{table:dataset}
\end{table*}
\section{UIL - Open Issues}
\label{sec:open}

Concerning the \textbf{RQ3}, \textit{What open issues still remain in UIL}, as detailed in the previous sections, researchers are proficiently working to provide more and more solutions to address feature extraction and model construction steps in the UIL framework in order to adequately face noisy, imperfect and largely unstructured data coming from different social networks. However, some open issues in this field still remain and they mainly concern: (i) data and datasets; (ii) evaluation; (iii) dynamic UIL; (iv) unsupervised models. 

Regarding the latter, it is useful to cite the new potential offered by the advent of Large Language Models (LLMs) \cite{vaswani2017attention, devlin2019bert,radford2019gpt2,brown2020gpt3} \ref{par:LLM}.
\paragraph{Data and datasets}
There is no established benchmark dataset for assessing and going beyond existing approaches. The number of publicly accessible datasets that have complete profiles, content, and network information is limited. However, existing datasets with partial features (like user names and network architecture) are available. A relevant issue related to the task concerns the ground truth. Finding user identity pairs that match across social media websites has become even more complex than before, especially when users make content private for the purpose. Also, getting a large dataset for research purposes is an open and challenging task. After the 2016 European GDPR, several concerns and limitations about user privacy were issued. Accessing user identity attributes and using them without violating the user's privacy is more challenging than before. From a practical point of view, OSN restrictions also limit the ability to crawl data via API. While some social networking websites offer APIs for adequate data access, they frequently impose rate limits and place restrictions on permission, which makes it challenging to collect data on a big scale.
\paragraph{Evaluation} The effectiveness of a UIL model could be assessed using a wide range of metrics. The review of the literature reveals that choosing a metric frequently depends on the kind of model used, making it difficult to consistently compare different models. The model, in turn, depends not only on the data sources but also on the specific application domains involved. It is still true that there is no definitive method for User Identity Linkage that is universally applicable. For instance, the models suitable vary based on whether you want a top-k matching or a perfect matching between pairs, and as a result, the metrics to utilize differ.
Furthermore, the substantial imbalance between matching and non-matching user identity pairs, which is a structural component of every dataset being worked on, has a significant impact on performance evaluation.

\paragraph{Dynamic UIL}
Furthermore, OSNs are always evolving in a dynamic manner. As time passes, profile, content, and network features for user identities continue to evolve and new links between friend users or links between the same person across different OSNs can be created. For this reason, the poor performance of a model for the UIL task could be motivated by a not-yet available online link but actually existing in the real world between the same person.
At the same time, several advancements have been accomplished in addressing the UIL task thanks to the advent of more effective deep learning architectures. Given the modern large pre-trained models \cite{vaswani2017attention} and the embedding-based models available \cite{mikolov2013distributed, mikolov2013efficient}, it is easier to involve user content on an OSN in the UIL task and compare embeddings of different users for tasks like link prediction and recommendation \cite{ting2016transfer, zheng2022you, li2017survey}. 
\paragraph{Unsupervised models}
In terms of unsupervised techniques, only three works had been examined up until 2016, and this field had been deemed to be understudied. The current review demonstrates that efforts to propose unsupervised methods have increased. We gathered five studies from after 2016 and three of them suggest novel unsupervised methods based on multi-dimensional embeddings and rich network data \cite{trung2020adaptive}, factoid embedding \cite{xie2018unsupervised}, and co-training algorithms \cite{zhong2018colink} that manipulate two independent models (attribute-based model and the relationship-based model), and makes them reinforce each other iteratively. However, although they exceed the state of the art of the previous unsupervised ones in performance, they do not outperform the state-of-the-art supervised, demonstrating that this area remains open for future research.
\paragraph{Potential Role of LLMs in accomplishing UIL task}
\label{par:LLM}
Large Language Models (LLMs) such as GPT-4, PaLM 2, Claude and LLaMA to cite the most powerful (according to ChatBot Arena's leaderboard\footnote{\url{https://chat.lmsys.org/} (2024-07-15)}):  offer advanced natural language processing capabilities that could significantly enhance User Identity Linkage (UIL) across social networks. These models excel at extracting and representing features from user-generated content, profiles, and behaviors, enabling the detection of similarities indicative of the same user across different platforms. By converting textual content into dense vector representations, LLMs create a unified embedding space that captures semantic nuances and contextual information, improving the accuracy of matching user profiles.

Additionally, LLMs can effectively handle entity resolution and disambiguation, identifying and differentiating between users with similar or identical names based on contextual cues. Their integration with graph-based methods, such as Graph Neural Networks (GNNs), further enhances UIL by leveraging both textual and structural data, thus providing a more comprehensive analysis. Moreover, LLMs address issues of data sparsity and heterogeneity through transfer learning and the imputation of missing data, ensuring better generalization across different platforms with varying data formats.

Privacy considerations are also crucial, and LLMs can be deployed in privacy-preserving frameworks like federated learning, allowing for decentralized data learning without compromising sensitive user information. They also facilitate anonymization and pseudonymization, enabling identity linkage without directly exposing personal identifiers. Overall, LLMs play a role in improving the accuracy, robustness, and scalability of UIL solutions, making them indispensable in the ongoing effort to link user identities across diverse and evolving social networks. 

To the best of our knowledge, there are currently no studies in the literature that utilize these language models to address the UIL task.
\section{Conclusion}
The process of tying together user accounts on different OSNs is challenging and attracted more and more research attention in the last two decades. The current work provides a comprehensive review of recent studies (from 2016 to the present) on User Identity Linkage (UIL) methods across online social networks by outlining various feature extraction strategies, algorithms, machine learning models, datasets, and evaluation metrics proposed by researchers working in this area. The proposed overview takes a pragmatic perspective to highlight the concrete possibilities for accomplishing this task depending on the type of available data. To this purpose, we offer a practical guide for other researchers in the field enriched with useful points of reference regarding algorithms, models, datasets, and evaluation metrics. The proposed excursus shows that several advances have been accomplished in addressing the UIL task thanks to the advent of more effective deep learning architectures. However, some issues still remain open and they mainly rely on the limited availability of benchmark datasets whose construction is even more complicated by the current social network access policies that reinforce privacy protection and reduce the possibility of accessing the data through API (see recent updates on Twitter APIs\footnote{\url{https://developer.twitter.com/en/docs/twitter-api/migrate/whats-new} (2024-09-10)}).
\label{sec:conclusion}

%\section*{Acknowledgment}

%%prova bibliography

\bibliographystyle{IEEEtran}
\bibliography{main}

\EOD

\end{document}